\documentclass[prl,preprint,tightenlines,%
superscriptaddress,showpacs,floatfix,nofootinbib,
]{revtex4}
\usepackage{graphics}
\usepackage{epsfig}
\usepackage{amssymb}

\newcommand{\ben}{\begin{displaymath}}
\newcommand{\een}{\end{displaymath}}
\newcommand{\be}{\begin{equation}}
\newcommand{\ee}{\end{equation}}
\newcommand{\bea}{\begin{eqnarray}}
\newcommand{\eea}{\end{eqnarray}}
\begin{document}
\title{Complex-mass renormalization in chiral effective field theory}
\author{D.~Djukanovic}
\affiliation{Institut f\"ur Kernphysik, Johannes
Gutenberg-Universit\"at, D-55099 Mainz, Germany}
\author{J.~Gegelia}
\affiliation{Institut f\"ur Kernphysik, Johannes
Gutenberg-Universit\"at, D-55099 Mainz, Germany} \affiliation{High
Energy Physics Institute of TSU, 0186 Tbilisi, Georgia}
\author{A.~Keller}
\affiliation{Institut f\"ur Kernphysik, Johannes
Gutenberg-Universit\"at, D-55099 Mainz, Germany}
\author{S.~Scherer}
\affiliation{Institut f\"ur Kernphysik, Johannes
Gutenberg-Universit\"at, D-55099 Mainz, Germany}
\date{25 February 2009}
\begin{abstract}
We consider a low-energy effective field theory of vector mesons
and Goldstone bosons using the complex-mass renormalization. As an
application we calculate the mass and the width of the $\rho$
meson.

\end{abstract}





\maketitle

\section{Introduction}

   The setting up of a consistent power counting scheme for chiral effective field
theories with heavy degrees of freedom is a 
non-trivial endeavor.
   For example, in baryon chiral perturbation theory the usual
power counting is not satisfied if the dimensional regularization is used
in combination with the minimal subtraction scheme
\cite{Gasser:1987rb}.
   The current solutions to this problem either involve the heavy-baryon approach
\cite{Jenkins:1990jv} or the use of a suitably chosen renormalization condition
\cite{Tang:1996ca,Becher:1999he,Gegelia:1999gf,Fuchs:2003qc}.
   Because the mass difference between the nucleon and the
$\Delta(1232)$ is small in comparison to the nucleon mass, the
$\Delta$ resonance can also be consistently included in the framework of effective
field theory
\cite{Hemmert:1997ye,Pascalutsa:2002pi,Bernard:2003xf,Hacker:2005fh}.

   On the other hand, the treatment of the $\rho$ meson is more
complicated.
   While the $\Delta$ resonance decays into a (heavy) nucleon and a (light) pion,
the main decay of the $\rho$ meson involves two pions with vanishing masses in the chiral limit.
   Therefore, for energies of the order of the $\rho$-meson mass, the loop diagrams
develop large imaginary parts.
   Unlike in the baryonic sector, these power-counting-violating contributions, being imaginary,
cannot be absorbed in the redefinition of the parameters of the Lagrangian as long as the
usual renormalization procedure is used.
   Despite this feature, the heavy-particle approach 
has been considered in
Refs.~\cite{Jenkins:1995vb,Bijnens:1996kg,Bijnens:1997ni,Bijnens:1997rv,Bijnens:1998di},
treating the vector mesons as heavy static matter fields.

    In Refs.~\cite{Fuchs:2003sh} and  \cite{Schindler:2003xv} we considered
the inclusion of virtual vector mesons in the framework of
(baryon) chiral perturbation theory for low-energy processes in which
the vector mesons cannot be generated explicitly.
   The present work extends the applicability of the chiral effective field theory to
the ''static'' properties of vector mesons.
   We tackle the power-counting problem by using the complex-mass renormalization
scheme
\cite{Stuart:1990,Denner:1999gp,Denner:2006ic,Denner:2005fg,Actis:2006rc,Actis:2008uh},
which is an extension of the on-mass-shell renormalization scheme
to unstable particles.
   As an application we consider the mass and
the width of the $\rho$ meson which are of particular interest
in the context of lattice extrapolations
\cite{Leinweber:1993yw,Leinweber:2001ac}.
   For a different approach to these problems using the infrared
regularization, see Refs.~\cite{Bruns:2004tj,Bruns:2008}.

\section{Lagrangian}

We start from the most general effective Lagrangian for $\rho$ and
$\omega$ mesons and pions in the parametrization of the model III
of Ref.~\cite{Ecker:1989yg}, where the $\rho$-vector fields
transform in-homogeneously under chiral transformations,
\begin{displaymath}
{\cal L}={\cal L}_{\pi}+{\cal L}_{\rho\pi}+{\cal L}_\omega
+{\cal L}_{\omega\rho\pi}+\cdots.
\end{displaymath}
   Displaying explicitly only those terms relevant for the calculations of
this work, the individual expressions read
\begin{eqnarray}
{\cal L}_\pi& = & \frac{F^2}{4}\,{\rm Tr} \left[\partial_\mu U
\left(\partial^\mu
U\right)^\dagger\right]+\frac{F^2\,M^2}{4}\,{\rm Tr} \left(
U^\dagger+U\right), \nonumber\\
{\cal L}_{\rho\pi}&=&
- \frac{1}{2}\,{\rm
Tr}\left(\rho_{\mu\nu}\rho^{\mu\nu}\right) + \left[ M_{\rho}^2 +
\frac{c_{x}\,M^2\,{\rm Tr} \left(
U^\dagger+U\right) }{4}\right]
{\rm Tr}\left[\left(
\rho^\mu-\frac{i\,\Gamma^\mu}{g}\right)\left(
\rho_{\mu}-\frac{i\,\Gamma_\mu}{g} \right)\right],\nonumber\\
{\cal L}_\omega&=&
 -\frac{1}{4}\left(
\partial_\mu\omega_{\nu}-\partial_\nu\omega_{\mu}\right)\left(
\partial^\mu\omega^\nu-\partial^\nu\omega^\mu\right)+\frac{M_{\omega}^2\,
\omega_{\mu}\omega^\mu}{2},\nonumber\\
{\cal L}_{\omega\rho\pi}&=&\frac{1}{2}\,
g_{\omega\rho\pi}\,\epsilon_{\mu\nu\alpha\beta}\, \omega^{\nu}\,
{\rm Tr}\left(\rho^{\alpha\beta} u^\mu
\right),\label{finallagrangian}
\end{eqnarray}
where
\begin{eqnarray}
U&=&u^2={\rm exp}\left(\frac{i\vec{\tau}\cdot\vec{\pi}}{F}\right),
\nonumber\\
\rho^\mu & = & \frac{\vec\tau\cdot\vec\rho\,^\mu}{2},\nonumber\\
\rho^{\mu\nu} & = &
\partial^\mu\rho^\nu-\partial^\nu\rho^\mu - i
g\left[\rho^\mu,\rho^\nu\right] \,,\nonumber\\
\Gamma_\mu &= & \frac{1}{2}\,\bigl[ u^\dagger\partial_\mu u+u
\partial_\mu u^\dagger
\bigr]\,, \nonumber
\\
u_\mu & = & i \left[ u^\dagger \partial_\mu u-u \partial_\mu
u^\dagger \right]. \label{somedefinitions}
\end{eqnarray}
In fact, at the beginning all the fields and parameters of Eqs.~(\ref{finallagrangian})
and (\ref{somedefinitions}) should be regarded as bare quantities which are usually indicated
by a subscript 0.
   However, in order to increase the readability of the expressions we have omitted this
index.
   In Eqs.~(\ref{finallagrangian}), $F$ denotes the pion-decay constant in the chiral
limit, $M^2$ is the lowest-order expression for the
squared pion mass, $M_\rho$ and $M_\omega$ refer to the bare $\rho$ and $\omega$ masses,
$g$, $c_x$, and $g_{\omega\rho\pi}$ are coupling constants.
   Demanding that the dimensionless and dimensionfull couplings are
independent, the consistency condition for the $\rho\pi\pi$
coupling \cite{Djukanovic:2004mm} leads to the KSFR relation
\cite{Kawarabayashi:1966kd,Riazuddin:sw}
\begin{eqnarray}
M_\rho^2 & = & 2\,g^2 F^2 \,.\label{M0}
\end{eqnarray}

\section{Renormalization and Power Counting}
   To perform the renormalization we use the standard procedure of
expressing the bare quantities (parameters and fields, now indicated by
a subscript 0) in terms of
renormalized ones, leading to the generation of counterterms.
   Below, we show explicitly only those which are
relevant for calculations of this work,
\begin{eqnarray}
\rho^\mu_0 & = & \sqrt{Z_\rho}\,\rho^\mu\,,
\nonumber\\
Z_\rho &=& 1 +\delta Z_\rho\,,\nonumber\\
M_{\rho,0} & = & M_R + \delta M_R \,,
\nonumber\\
c_{x,0}  & = & c_x+\delta c_x.
\label{renormparameters}
\end{eqnarray}
We apply the complex-mass renormalization scheme
\cite{Stuart:1990,Denner:1999gp,Denner:2006ic,Denner:2005fg,Actis:2006rc,Actis:2008uh}
and choose $M_R^2=(M_\chi - i\, \Gamma_\chi/2)^2$ as the pole of
the $\rho$-meson propagator in the chiral limit.
   The pole mass and the width of the $\rho$ meson
in the chiral limit are denoted by $M_\chi$ and
$\Gamma_\chi$, respectively.
   Both are input parameters in our approach.
   We include $M_R$ in the propagator and the counterterms are treated
perturbatively.
   In the complex-mass renormalization scheme, the counterterms are also complex
quantities.


   Since the $\rho$ mass will not be treated as a small quantity, the
presence of large external four-momenta, e.g., in terms of the zeroth component,
leads to a considerable complication regarding the power counting of loop diagrams.
   To assign a chiral order to a given diagram it is first necessary to investigate
all possibilities how the external momenta could flow through the
internal lines of that diagram.
   Next, when assigning powers to propagators and vertices, one needs to determine
the chiral order for a given flow of external momenta.
   Finally, the smallest order resulting from the various assignments
is defined as the chiral order of the given diagram.

   The power counting rules are as follows. Let $q$ collectively stand for a small quantity
such as the pion mass.
   A pion propagator counts as ${\cal O}(q^{-2})$ if it does not
carry large external momenta and as ${\cal O}(q^{0})$ if it does.
   On the other hand, a vector-meson propagator counts as ${\cal O}(q^{0})$
if it does not carry large external momenta and as ${\cal O}(q^{-1})$ if it does.
   The pion mass counts as ${\cal O}(q^{1})$, the vector-meson mass as ${\cal O}(q^{0})$,
and the width as ${\cal O}(q^{1})$.
   Vertices generated by the effective Lagrangian of Goldstone bosons ${\cal
L}_\pi^{(n)}$ count as ${\cal O}(q^n)$.
   Derivatives acting on heavy vector mesons, which cannot be eliminated by field
redefinitions, count as ${\cal O}(q^0)$. The contributions of
vector meson loops can be absorbed systematically in the
renormalization of the parameters of the effective Lagrangian.
Therefore, such loop diagrams need not be included for energies
much lower than twice the vector-meson mass.

\section{Evaluation of the Two-Point Function}
   The mass and width of the $\rho$ meson are extracted from the complex pole of
the two-point function.
   The undressed propagator of the vector meson reads
\begin{equation}
i\,S^{a b}_{0\, \mu\nu}(p) = -i
\,\delta^{ab}\frac{g_{\mu\nu}-\frac{p_\mu p_\nu}{M_R^2}}{p^2 -
M_R^2+i\,0^+}\,,\label{bareprop}
\end{equation}
with complex $M_R^2$.
   We parameterize the sum of all one-particle-irreducible diagrams
contributing to the two-point function as
\begin{equation}
i\,\Pi^{ab}_{\mu\nu}(p)=i\,\delta^{ab}\,\left[ g_{\mu\nu}\Pi_{1}
(p^2)+p_\mu p_\nu\,\Pi_2 (p^2)\right]\,. \label{VSEpar}
\end{equation}
The dressed propagator, expressed in terms of the self energy, has
the form
\begin{equation}
i\, S^{a b}_{\mu\nu}(p) = -i
\,\delta^{ab}\,\frac{g_{\mu\nu}-p_\mu p_\nu\frac{1+
\Pi_2(p^2)}{M_R^2+\Pi_1(p^2)+p^2 \Pi_2(p^2)}}{p^2 -
M_R^2-\Pi_1(p^2)+i\,0^+}\,.\label{dressedprop}
\end{equation}
   The pole of the propagator is found as the (complex) solution to
the following equation:
\begin{equation}
z- M_R^2-\Pi_1(z)=0\,. \label{poleequation1}
\end{equation}
In the vicinity of the pole $z$, the dressed propagator can be
written as
\begin{equation}
i\, S^{a b}_{\mu\nu}(p) = -i \delta^{ab}\left[\frac{Z^r_\rho
\left(g_{\mu\nu}-\frac{p_\mu p_\nu}{z}\right)}{p^2
-z+i\,0^+}+R\right]\,,\label{dressedpropnearpole}
\end{equation}
where
$$
Z^r_\rho=\frac{1}{1- \Pi_1'(z)}\,,
$$
and $R$ stands for the non-pole part.
   The counterterms $\delta M_R$ and $\delta Z_\rho$ are fixed by requiring
that, in the chiral limit, $M_R^2$ is the pole of the dressed
propagator and that the residue $Z_\rho^r$ is equal to one.

\begin{figure}
\epsfig{file=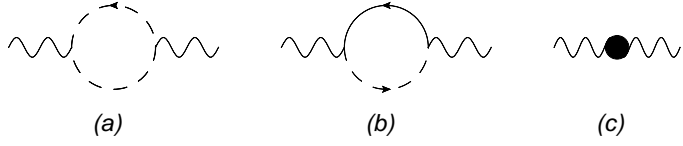, width=9truecm}
\caption[]{\label{mnloops:fig} One-loop contributions to the
$\rho$-meson self-energy at ${\cal O}(q^3)$. The dashed, solid,
and wiggly lines correspond to the pion, the $\omega$ meson, and
the $\rho$ meson, respectively.}
\end{figure}

   The solution to Eq.~(\ref{poleequation1}) can be found
perturbatively as an expansion
\begin{equation}
z=z^{(0)}+ z^{(1)}+z^{(2)}+\cdots\,,\label{Mpoleloopexpansion}
\end{equation}
where the superscripts $(i)$ denote the $i$th-loop order.
   Each of these terms can be expanded in small quantities in the chiral expansion.
   Up to and including third chiral order, the tree-order result for $\Pi_1$ is
\begin{equation}
\Pi_1^{(0)} = c_x M^2\,.\label{treerhoSE}
\end{equation}
   At tree order, the pole obtained from Eq.~(\ref{poleequation1}) reads
\begin{equation}
z^{(0)}= M_R^2+c_x M^2\,. \label{poleequation3}
\end{equation}
   The one-loop contributions to the vector self-energy up to ${\cal O}(q^3)$ are shown in
Fig.~\ref{mnloops:fig}.
   The contributions of diagrams (a) and (b) to $\Pi_1$
are given by
\begin{eqnarray}
D_{a} & = & -\frac{g^2 \mu ^{4-n}   \left[2 \,I_{M }-\left(p ^2-4
M^2\right) \,I_{MM}\right]}{n-1}\,, \nonumber\label{diag24} \\
D_{b} & = & \frac{(n-2)\,g_{\omega \rho \pi}^2\,\mu^{4-n}}{4\,
(n-1)}\,\left[ M^4\,I_{MM_{\omega}}
 -\left(2\,I_{MM_{\omega}}
M_\omega^2+I_M-I_{M_\omega}+2\,I_{MM_\omega} p^{2}\right)
M^2\right.\nonumber\\
&& \left.+I_{MM_{\omega}} p^{2 }+M_\omega^2 \left(I_{MM_{\omega}}
M_\omega^2+I_M-I_{M_\omega}\right)
-\left(2 I_{MM_{\omega}} M_\omega^2+I_M+I_{M_\omega}\right)
p^{2}\right]\,, \label{diag25}
\end{eqnarray}
where the loop integrals are defined as
\begin{eqnarray}
I_{m_1m_2} & = & \frac{i}{(2 \pi)^{n}}\,\int
\frac{d^nk}{\left[k^2-m_1^2+i\,0^+\right]\left[(p+k)^2-m_2^2+i\,0^+\right]}\,,\nonumber\\
I_m & = & \frac{i}{(2 \pi)^{n}}\,\int \frac{
d^nk}{k^2-m^2+i\,0^+}\,, \label{oneandtwoPF}
\end{eqnarray}
with $n$ the space-time dimension and $p$ the four-momentum of the
vector meson.

Due to the large momenta flowing through the $\rho\pi\pi$ vertex
in diagram (a), this vertex should, in principle, count as ${\cal
O}(q^0)$.
   However, its large component is proportional to $p^\mu$ and, thus,
does not contribute to $\Pi_1$.
   Therefore, the $\rho\pi\pi$ vertex actually contributes as ${\cal
O}(q^1)$. Hence, diagram (a) contributes to $\Pi_1$ starting at
${\cal O}(q^4)$, which is beyond the accuracy of our calculation.
Diagram (c) contains the contributions of the counterterms.

   Diagram (a) contains a power-counting-violating imaginary part
(which is proportional to the $\rho$-meson mass for an
''on-shell'' resonance and hence does not vanish in the chiral
limit).
   It is impossible to cancel this imaginary part by contributions of
counterterms unless we use the complex-mass renormalization
scheme, where the counterterm contributions become complex
quantities.
   It is this new feature which makes a crucial difference and allows
one to solve the power-counting problem for the ''on-shell''
$\rho$ meson. In diagram (b) we take $M_\omega = M_R$ which is a
good approximation for the purposes of this work.

We fix the counterterms contributing to the pole of the
$\rho$-meson propagator such that the pole at chiral limit stays
at $M_R^2$. This gives:
\begin{eqnarray}
\delta M_R & = & -\frac{1}{3}\, g^2 M_R \,\lambda + \frac{g^2\,
M_R \left(-3 \ln \frac{M_R^2}{\mu^2 }+3\, i\, \pi +5\right)}{288\,
\pi ^2} + \frac{1}{3} g_{\omega\rho\pi}^2 M_R^3
\lambda+\frac{g_{\omega\rho\pi}^2 M_R^3 (3 \ln
\frac{M_R^2}{\mu^2}+1)}{288 \pi ^2}\,,\nonumber\\
\delta c_x & = & 4\,g^2 \,\lambda - \frac{g^2 \left( 1 - \ln
\frac{M_R^2}{\mu^2 }+ i \pi \right)}{8 \pi ^2} +
g_{\omega\rho\pi}^2 M_R^2 \lambda - \frac{g_{\omega\rho\pi}^2
M_R^2 \left(1- \ln\frac{M_R^2}{\mu^2}\right)}{32 \pi ^2}\,,
\label{deltaZ}
\end{eqnarray}
where
\begin{equation}
\lambda =
\frac{1}{16\,\pi^2}\left\{\frac{1}{n-4}-\frac{1}{2}\,\left[\ln(4
\pi)+\Gamma'(1)+1\right]\right\}\,. \label{lambdadef}
\end{equation}
The contributions  of diagrams (a), (b) and (c) to the pole,
expanded up to ${\cal O}(q^4)$, read
\begin{equation}
z^{(1)}=\frac{g^2M^4 }{16\,\pi^2 \,M_R^2}\left(3 -2\, \ln
\frac{M^2}{M_R^2}-2\,i\,\pi \right)
-\frac{g_{\omega\rho\pi}^2 M^3 M_\chi}{24 \pi}
-\frac{g_{\omega\rho\pi}^2 M^4 \left(\ln
\frac{M^2}{M_\chi^2}-1\right)}{32 \pi ^2}
+ \frac{i g_{\omega \rho \pi}^2 \,M^3 \Gamma_\chi }{48 \pi }\,.
\label{deltaM1}
\end{equation}
As is seen from Eq.~(\ref{deltaM1}), the contribution of diagram
(c) is indeed of ${\cal O}(q^4)$ within the complex-mass
renormalization scheme.

Using the renormalized version of Eq.~(\ref{M0}), i.e.,  $M_R^2=2
g^2 F^2$, to eliminate $g^2$ from Eq.~(\ref{deltaM1}), we obtain
for the pole mass and the width of the $\rho$ meson to ${\cal
O}(q^4)$
\begin{eqnarray}
M_\rho^2  & = & M_\chi^2 +c_x M^2
-\frac{g_{\omega\rho\pi}^2 M^3 M_\chi}{24 \pi }+
\frac{M^4}{32\pi^2 F^2} \left(3 -2\, \ln
\frac{M^2}{M_\chi^2}\right) -\frac{g_{\omega\rho\pi}^2 M^4
\left(\ln \frac{M^2}{M_\chi^2}-1\right)}{32 \pi ^2}\,,
\nonumber\\
\label{phmass}\\
\Gamma & = & \Gamma_\chi +\frac{\Gamma_\chi ^3}{8
M_\chi^2}-\frac{c_x \Gamma_\chi  M^2}{2 M_\chi^2}
-\frac{g_{\omega\rho\pi}^2 M^3 \Gamma_\chi}{48 \pi \,M_\chi}
+\frac{ M^4}{16\,\pi \,F^2 M_\chi}\,. \label{phwidth}
\end{eqnarray}
The non-analytic terms of Eq.~(\ref{phmass}) agree with the results of
Ref.~\cite{Leinweber:2001ac}.
   Note that both mass $M_\chi$  and width $\Gamma_\chi$ in the chiral
limit are input parameters in our approach.

To estimate the numerical values of
contributions of different orders we substitute
$$
F=0.092 \,{\rm GeV},\quad M=0.139 \,{\rm GeV}\,,\quad g_{\omega\rho\pi} = 16 \,{\rm
GeV^{-1}},\quad M_\chi\approx M_\rho=0.78\,{\rm GeV}
$$
and obtain in units of GeV$^2$ and GeV, respectively,
\begin{eqnarray}
M_\rho^2 & = & M_\chi^2+0.019 \,c_x - 0.0071
 + 0.0014
+0.0013\,,\nonumber\\
\Gamma & \approx & \Gamma_\chi + 0.21\,\Gamma_\chi^3-0.016\,c_x
\Gamma_\chi -0.0058\,\Gamma_\chi + 0.0011\,. \label{numestimate}
\end{eqnarray}
For pion masses larger than $M_\rho/2$ the $\rho$ meson becomes
a stable particle. For such values of the pion mass the series of
Eq.~(\ref{phwidth}) should diverge.

\section{Conclusions}
    To summarize, we have considered an effective field theory of vector
mesons interacting with Goldstone bosons using the complex-mass
renormalization scheme. A systematic power counting (at least for
the ''static'' properties of the vector mesons) emerging within
this scheme allows one to calculate the physical quantities in
powers of small parameters. As an application we have calculated
the pole mass and the width of the $\rho$ meson  which are of
particular interest in the context of lattice extrapolations
\cite{Leinweber:1993yw,Leinweber:2001ac}. In the isospin-symmetric
limit, we calculated these quantities to ${\cal O}(q^3)$ in terms
of the light quark mass and the width of the vector meson in the
chiral limit. {To estimate the contributions of higher orders we
also retained ${\cal O}(q^4)$ terms of ${\cal O}(q^3)$ diagrams.

\medskip

\acknowledgments

This work was supported by the Deutsche Forschungsgemeinschaft
(SFB 443).

\end{document}